# Phonons and Colossal Thermal Expansion Behavior of $Ag_3Co(CN)_6$ and $Ag_3Fe(CN)_6$

R Mittal[1], M. Zbiri[2], H. Schober[2,3], S. N. Achary[4], A. K. Tyagi[4] and S. L. Chaplot[1]

[1]*Solid State Physics Division, Bhabha Atomic Research Centre, Trombay, Mumbai 400 085, India*

[2]*Institut Laue-Langevin, BP 156, 38042 Grenoble Cedex 9, France*

[3]*Université Joseph Fourier, UFR de Physique, 38041, Grenoble Cedex 9, France*

[4]*Chemistry Division, Bhabha Atomic Research Centre, Trombay, Mumbai, 400085, India*

Recently colossal positive volume thermal expansion has been found in the framework compounds $Ag_3Co(CN)_6$ and $Ag_3Fe(CN)_6$. Phonon spectra have been measured using the inelastic neutron scattering technique as a function of temperature and pressure. The data has been analyzed using *ab-initio* calculations. We find that the bonding is very similar in both compounds. At ambient pressure modes in the intermediate frequency part of the vibrational spectra in the Co compound are shifted to slightly higher energies as compared to the Fe compound. The temperature dependence of the phonon spectra gives evidence for large explicit anharmonic contribution to the total anharmonicity for low-energy modes below 5 meV. We found that modes are mainly affected by the change in the size of unit cell, which in turn changes the bond lengths and vibrational frequencies. Thermal expansion has been calculated via the volume dependence of phonon spectra. Our analysis indicates that Ag phonon modes in the energy range from 2 to 5 meV are strongly anharmonic and major contributors to thermal expansion in both compounds. The application of pressure hardens the low-energy part of the phonon spectra involving Ag vibrations and confirms the highly anharmonic nature of these modes.

Compounds that contract upon heating over a certain temperature range are the exception. They, however, find extremely useful technical application in composites. Compensating the usual thermal expansion of ordinary materials contracting compounds allow tailoring the thermal expansion behavior of the ensemble [1]. The technological relevance of these compounds and their composites principally derives from their ability to withstand thermal shock without any damage. The negative thermal expansion (NTE) phenomenon typically originates [2-6] from the presence of low-energy anharmonic vibrations.

Structural flexibility plays a key role in the physical properties of framework materials. NTE has been found [7,8] in framework oxides (-27 $\times$ 10$^{-6}$ K$^{-1}$ for $ZrW_2O_8$ at 300 K) with M-O-M (M = metal) linkages. The Zr-O-W angle in these compounds is sensitive to changes in temperature, pressure and composition. Further, the NTE coefficients of compounds with M–CN–M linkages between two neighboring tetrahedral units, instead of a single atom, was found to reach a much higher value of -51$\times$ 10$^{-6}$ K$^{-1}$ for $Zn(CN)_2$ [9] at 300 K. The linear CN unit acts like a stick adding additional flexibility to the network. This may explain why NTE appears to be inherently more common amongst cyanides than amongst oxide-containing frameworks, and also why the magnitude of the NTE coefficient can be much larger.

The concept of increasing framework flexibility has led to the discovery [10, 11] of colossal positive and negative thermal expansion in $Ag_3Co(CN)_6$ and $Ag_3Fe(CN)_6$. The crystal lattice expands remarkably along the a and b crystal axes (~ 120$\times$10$^{-6}$ K$^{-1}$), while it contracts equally strongly (~ -110$\times$10$^{-6}$ K$^{-1}$) along the c axis. The structure thus expands and contracts at rates more than ten times larger than what is found in normal materials. The covalent framework structure of this material consists of Co–CN–Ag–CN–Co linkages. It features alternating layers of octahedral [Co(CN)6]$^{3-}$ and Ag$^+$ cations (Fig. 1). At ambient pressure $Ag_3Co(CN)_6$ is reported to decompose [10] at 500 K. High pressure diffraction measurements [12] show a transition from a trigonal to a monoclinic structure at very low pressures of 0.19 GPa at 300 K. The high-pressure phase is known to be 16% more dense as compared to the ambient pressure phase. High-pressure Raman measurements indicate [13] that $Ag_3Co(CN)_6$ becomes amorphous around 13 GPa.

As far as we know the phonon spectrum have not been measured for $Ag_3Co(CN)_6$. However, a reciprocal-space analysis of neutron diffraction data [10] has been used to obtain average phonon energies of vibrations of Ag, Co, and CN as a function of temperature. The average phonon energy of Ag vibrations has been reported to decrease noticeably with increasing temperature, whereas there is very little change in the energies of Co and CN vibrations. The anharmonicity of Ag vibrations appears to drive the colossal thermal expansion of the flexible lattice of $Ag_3Co(CN)_6$.

The origin of anomalous thermal expansion behavior has been widely studied [3-6,14-16] in many compounds. The low energy vibrational modes play an important role in the observed anomalous behavior of the lattice parameters. Recently we have reported [14,15] inelastic neutron scattering and lattice dynamical calculations for $Zn(CN)_2$ and $Ni(CN)_2$ to understand anomalous thermal expansion behavior. Here we report our studies on $Ag_3Co(CN)_6$ and $Ag_3Fe(CN)_6$ to understand the anharmonicity of phonons responsible for anomalous large thermal expansion coefficients. We have measured the temperature dependence of the phonon spectra in $Ag_3Co(CN)_6$ and $Ag_3Fe(CN)_6$. Further, phonon spectra has also been measured in the high-pressure phase of $Ag_3Co(CN)_6$. The analysis of the experiments is performed using *ab-initio* lattice dynamical calculations.

## II. EXPERIMENTAL

Both $Ag_3Fe(CN)_6$ and $Ag_3Co(CN)_6$ were prepared by precipitation from the aqueous solution of $K_3Fe(CN)_6$ or $K_3Co(CN)_6$ and $AgNO_3$ in a similar procedure reported by Goodwin et al. [10]. About 6.47 g of $K_3Fe(CN)_6$ and 10.23 g of $AgNO_3$ were dissolved separately in 60 ml of distilled water and stirred well to make clear solutions. The $AgNO_3$ solution was added slowly to the solution of $K_3Fe(CN)_6$ with constant stirring. Dark orange red colored precipitates of $Ag_3Fe(CN)_6$ formed instantaneously at room temperature. The suspension is kept stirring for about 2 h at ambient temperature and then filtered. The residue obtained was washed repeated with distilled water and dried at 50°C for 5h. About 3.78 g of $K_3Co(CN)_6$ and 5.12 g of $AgNO_3$ were dissolved separately in 30 ml of distilled water and stirred well to make clear solutions. The $AgNO_3$ solution was added slowly to the solution of $K_3Co(CN)_6$ with constant stirring. White precipitates of $Ag_3Co(CN)_6$ formed instantaneously at room temperature. Further the total content was heated to about 100 °C with constant stirring for about 2 h and then allowed to cool to ambient temperature. The precipitate was separated by filtering and the residue was washed repeatedly by hot distilled water and then cold water. The product was dried at 50°C for about 12 h. The final product $Ag_3Co(CN)_6$ (white powder) and $Ag_3Fe(CN)_6$ (dark green) were characterized powder x-ray diffraction pattern. The powder X-ray diffraction pattern was recorded in the two theta range of 10-90° using CuKα x-ray. The observed powder x-ray diffraction data was analyzed by Rietveld refinement method. The refined unit cell parameters are: a = 7.008(4) and c =

7.225(4) Å, V = 307.3(4) Å$^3$ (Space group: $P\bar{3}1m$) for Ag$_3$Fe(CN)$_6$ and a = 7.025(1) and c = 7.102(1) Å (Space group: $P\bar{3}1m$) for Ag$_3$Co(CN)$_6$.

The temperature-dependent inelastic neutron scattering experiments were performed using the IN6 cold neutron time-of-flight time-focusing spectrometer at the Institut Laue Langevin (ILL), France. The measurements were performed on about 8 grams of polycrystalline samples of Ag$_3$Co(CN)$_6$ and Ag$_3$Fe(CN)$_6$. The high-resolution data was collected, at several temperatures from 160 K to 300 K, in the neutron-energy gain inelastic focusing mode with incident neutron wavelength of 4.14 Å (4.77 meV). In this configuration the energy resolution of the spectrometer is 0.2 meV at the elastic position.

The high pressure measurements were carried out with incident neutron wavelength of 5.12 Å (3.12 meV) was chosen for the measurements, which led to a resolution of 0.07 meV at the elastic line. The sample was compressed using argon gas in a pressure cell available at ILL. The inelastic neutron scattering spectrum have been measured at 200 K for a polycrystalline sample of Ag$_3$Co(CN)$_6$ at ambient pressure, 0.3, 1.9 and 2.8 kbar. The inelastic neutron scattering signal is corrected for the contributions from argon at the respective pressures and for the empty cell. A standard vanadium sample was used to calibrate the detectors.

In the incoherent one-phonon approximation the measured scattering function $S(Q,E)$, as observed in the neutron experiments, is related to the phonon density of states [17] as follows:

$$g^{(n)}(E) = A \left\langle \frac{e^{2W_k(Q)}}{Q^2} \frac{E}{n(E,T) + \frac{1}{2} \pm \frac{1}{2}} S(Q,E) \right\rangle \quad (1)$$

$$g^n(E) = B \sum_k \left\{ \frac{4\pi b_k^2}{m_k} \right\} g_k(E) \quad (2)$$

where the + or − signs correspond to energy loss or gain of the neutrons, respectively, and $n(E,T) = \left[\exp(E/k_BT) - 1\right]^{-1}$. $A$ and $B$ are normalization constants and $b_k$, $m_k$, and $g_k(E)$ are, respectively, the neutron scattering length, mass, and partial density of states of the $k^{th}$ atom in the unit cell. The quantity within < ---- > represents an appropriate average over all $Q$ values at a given energy. $2W(Q)$ is the Debye-Waller factor. The weighting factors $\frac{4\pi b_k^2}{m_k}$ for each atom type in the units of barns/amu are: Ag: 0.046; Co: 0.095; Fe: 0.208; C: 0.462 and N: 0.822 calculated from the neutron scattering lengths found in Ref. [18].

## III. LATTICE DYNAMICAL CALCULATIONS

Ab-initio calculations were performed using the projector-augmented wave (PAW) formalism [19] of the Kohn-Sham DFT [20, 21] at the generalized gradient approximation level (GGA), implemented in the Vienna ab initio simulation package (VASP) [22, 23]. The GGA was formulated by the Perdew-Burke-Ernzerhof (PBE) [24, 25] density functional. The Gaussian broadening technique was adopted and all results are well converged with respect to k-mesh and energy cutoff for the plane wave expansion. Partial geometry optimization, where the lattice parameters were kept fixed, was carried out on the experimentally refined $Ag_3[Co(CN)_6]$ [10] and $Ag_3[Fe(CN)_6]$ [11] structures containing four crystallographically inequivalent atoms, and having the hexagonal $P\bar{3}1m$ space group (162 $[D_{3d}^1]$). In the lattice dynamics calculations, in order to determine all force constants, the super cell approach was used [26]. An orthorhombic super cell was constructed from the partially relaxed structure containing 8 formula-units (128 atoms). Total energies and Hellmann-Feynman forces were calculated for 42 structures resulting from individual displacements of the symmetry inequivalent atoms in the super cell, along the three inequivalent cartesian directions (±x, ±y and ±z). PDS, PDR and Raman frequencies were extracted in subsequent calculations using the Phonon software [27].

## IV. RESULTS AND DISCUSSION

**A. Phonon Spectra of $Ag_3Co(CN)_6$ and $Ag_3Fe(CN)_6$**

The spectra from $Ag_3Co(CN)_6$ and $Ag_3Fe(CN)_6$ are shown in Figs. 2 and 3. Measurements were carried out at 160, 190, 220, 250 and 300 K. Please note that due to the use of the energy gain mode the frequency window is limited to below 80 meV at lower temperatures. This is, however sufficient to observe all external modes. The spectra show several vibrational bands centered on 3, 13, 17, 22, 45, 55 and 70 meV (Figs. 2 and 3). The ab-initio calculations (Figs. 2 and 3) reproduce very well the positions of the bands while there are slight differences with respect to the intensities.

The temperature dependence of the overall spectra is rather weak for the interval considered. For thermal expansion the low-frequency part, as shown in Fig. 4 for 160 K and 300 K, is particularly interesting. The low-energy modes are found to harden anomalously with increasing temperature. Below 5 meV, the total anharmonicity, sum of implicit and explicit contributions, $\left(\frac{1}{E_i}\frac{dE_i}{dT}\bigg|_P\right)$, of the low-energy modes is thus found (Fig. 4) to be positive. The implicit anharmonicity, i.e. the volume dependence of the phonon spectra, should result in a decrease of phonon frequencies for compounds with colossal thermal expansion behavior. The hardening of modes with increase of temperature gives us, therefore, evidence for the large explicit anharmonicity, i.e. changes in phonon frequencies due to large thermal amplitude of atoms, reflecting the nature of phonons in these compounds.

The phonon spectra of both compounds are very similar (Fig. 5). In particular, the cut-off energy for the external modes in both compounds is at about 80 meV. The main difference resides in a more smeared out character for the modes in the Fe compound above 40 meV. The mass effect upon substitution of Co (58.93 amu) by Fe (55.84 amu) would only slightly shift the modes associated with the Co/Fe atoms. The similarity of the spectra thus indicates that the strength and character of bonding is nearly the same for both systems. The calculated partial density of states (Fig. 6) shows that Ag atoms mainly contribute in the low-energy modes up to 10 meV. The vibrations due to C and N span the entire energy spectrum up to 280 meV. Co (58.93 amu) and Fe (55.84 amu) are found in the same range of up to 75 meV.

The temperature-dependence of the Bose factor corrected scattering function S(Q,E) is plotted on Figure 7. The figure clearly shows the presence of flat dispersion surfaces at 2.5 and 4.5 meV. These produce the two-peak structure in the spectra (Figure 4). The temperature dependence of S(Q,E) plots also indicate that energies of the flat acoustic modes do not change significantly as a function of temperature, which is in agreement with the phonon spectra shown in Figs. 2 and 3.

The calculated phonon dispersion curves for $Ag_3Co(CN)_6$ and $Ag_3Fe(CN)_6$ are shown in Fig. 8. The flat phonon dispersion sheet of the two lowest energy acoustic modes near the zone boundary at about 3 meV give rise to the observed first peak in the density of states. The second low-frequency peak at about 4.5 meV corresponds to flat optic modes. We also notice flat phonon dispersion sheets in the entire Brillouin zone at relatively high energies of about 40, 52, 58 and 75 meV (Fig. 9) providing the other well isolated bands in the phonon density of states (Fig. 3). The dispersion relation is found to be quite similar in both the compounds.

The temperature dependence of the average value of phonon energies of Ag, Co, and CN species in $Ag_3Co(CN)_6$ have been obtained [10] from the reciprocal-space analysis of diffraction data. We have directly calculated the phonon spectra of these compounds for the structural data [10] at various temperatures. The partial density of states corresponding to the structural data at various temperatures has been used to obtain the mean vibrational energy of atoms. The comparison of the mean partial phonon frequencies of various atoms as obtained from reciprocal-space analysis of neutron diffraction data and our ab-inito calculations are shown in Fig. 9. We find large deviations in the results. This demonstrates that in case of very complex systems special care has to be taken when extracting information on phonon spectra in an indirect way from structural data. The analysis of diffraction data equally indicates a decrease of the average phonon energy of Ag with increasing temperature, whereas those for Co and CN sub-lattice are predicted to change little. Our measurements show that the average energies of all the atoms remain nearly constant. Further we have also calculated average energies of atoms in $Ag_3Fe(CN)_6$, which are found to be slightly lower as compared to those in $Ag_3Co(CN)_6$. This may be attributed to the slightly higher unit cell volume in the Fe compound as compared to that of the Co compound.

The structures of $Ag_3Co(CN)_6$ and $Ag_3Fe(CN)_6$ yield 48 phonon modes for each wave vector. The Raman and infrared spectra of these compounds have not been previously reported. The calculated zone centre modes for both compounds are given in Table I. It can be seen that the range of external modes is roughly same for both systems. However, the C≡N stretching modes of the Co compounds are shifted to slightly higher energies as compared to the Fe compounds due to the slightly lower unit cell volume in the former compound.

**B. High-Pressure Inelastic Neutron Scattering Measurements on $Ag_3Co(CN)_6$**

$Ag_3Co(CN)_6$ is known to undergo a structural phase transition [12] at very low pressure of 1.9 kbar at 300 K. We have measured (Fig. 10) phonon spectra of $Ag_3Co(CN)_6$ at ambient pressure, 0.3, 1.9 and 2.8 kbar, at 200 K. The large absorption from the Ag and Co atoms as well as from the high-pressure cell has enabled us to measure phonon spectra only up to 10 meV. We find that at 1.9 kbar the low-energy part of the phonon spectra up to 5 meV is most significantly affected by the application of pressure. The height of the peaks in the phonon spectra is found to reduce and the modes around 2.5 and 4.5 meV shift to higher energies. As discussed above the low-energy part of the phonon spectra features mainly contributions from Ag atoms and these modes are strongly anharmonic. The measurements provide indirect evidence that atomic motions in the Ag sub-lattice may be responsible for the observed phase transition in $Ag_3Co(CN)_6$ at very low pressure. This is in agreement with the diffraction measurements [12] where a weak argentophilic Ag–Ag interaction is found to drive the triclinic to monoclinic phase transition.

**C. Thermal Expansion Behaviour of $Ag_3Co(CN)_6$ and $Ag_3Fe(CN)_6$**

The calculation of the thermal expansion is carried out in the quasi-harmonic approximation for which each phonon mode contributes to the volume thermal expansion coefficient [28] given by: $\alpha_V = \frac{1}{BV}\sum_i \Gamma_i C_{Vi}(T)$, with $\Gamma_i$ $(=-\partial \ln E_i/\partial \ln V)$ and $C_{Vi}$ the mode Grüneisen parameter and specific heat of the $i^{th}$ vibrational state of the crystal, respectively. In order to estimate theoretically the Grüneisen parameters, we have calculated the phonon density of states for partially relaxed structures by keeping the experimental lattice parameters corresponding to 16 (16) K and 196 (194) K for $Ag_3Co(CN)_6$ ($Ag_3Fe(CN)_6$). The calculated Grüneisen parameter, $\Gamma(E)$ as shown in Fig. 11 is averaged over all phonons of energy E in the Brillouin zone. We find that low-energy modes around 2 meV have very large positive $\Gamma(E)$. The calculated temperature dependence of the thermal expansion behavior as shown in Fig. 12 is in reasonable agreement with the experimental data. However, qualitatively the calculations are useful for understanding the colossal thermal expansion behavior of $Ag_3M(CN)_6$ (M=Co, Fe).

The calculated contribution of phonons of energy $E$ to the volume thermal expansion coefficient ($\alpha_V$) as a function of energy at 300 K (Fig. 13) shows that the maximum contribution to $\alpha_V$ is from modes between 2 and 5 meV. The calculated partial density of states identifies these modes as mainly involving Ag atom vibrations. It is the anharmonicity of Ag vibrations that drives the thermal expansion of these flexible lattices.

The crystal structure of $Ag_3Co(CN)_6$ consists of Co-CN-Ag-NC-Co linkage. The trigonal unit cell has $CoC_6$ octahedral units. The C atoms are covalently connected to the N atoms. Further the two alternating layers of $[Co(CN)_6]^{3-}$ ions are connected by Ag atoms. In order to understand the nature of phonons responsible for anomalous thermal expansion we have calculated mean squared displacements (Fig. 14) of various atoms, $<u^2>$, arising from all phonons of energy E in the Brillouin zone. The calculated partial density of states (Fig. 5) has been used for this calculation. Equal amplitude for all the atoms up to about 1.5 meV identifies the acoustic region. The larger amplitude of C atoms as compared to Co/Fe atoms in the 2 to 5 meV energy indicates librational motion of $CoC_6$ octahedra. The large amplitude of Ag and N atoms indicates their translational motion.

The thermal expansion behavior for isostructural $D_3[Co(CN)_6]$ has been reported [11] in literature. The comparison of all the three compounds show that there is colossal thermal expansion behavior in $Ag_3M(CN)_6$ (M=Co, Fe), while the thermal expansion coefficient in $D_3[Co(CN)_6]$ is an order of magnitude less then $Ag_3M(CN)_6$. The experimental data [11] clearly show that chemical variation at the octahedral transition metal (M=Co, Fe) site does not change the thermal expansion behavior, while the substitution of Ag by D has a large impact. As discussed above we find that our calculated thermal expansion coefficient is almost the same in both the $Ag_3M(CN)_6$ (M=Co, Fe) compounds. It is clear that it is the anharmonicity of Ag (107.8) or D (2 amu) vibrations that makes a difference in thermal expansion behavior of these compounds. The mass difference of Ag and D atoms would shift the modes of D atoms to higher energies in $D_3[Co(CN)_6]$ as compared to modes of Ag in $Ag_3M(CN)_6$, which in turn lowers the entropy and thermal expansion coefficient.

## V. CONCLUSIONS

The colossal thermal expansion behavior in $Ag_3M(CN)_6$ (M=Co, Fe) has been investigated using inelastic neutron spectroscopy and ab-initio calculations to probe directly the phonon spectra as a function of pressure and temperature. The measured spectra are very well reproduced by the calculations. The intermediate and high-energy parts of the vibrational spectra for the Co compound are found to shift to slightly higher energies in comparison with the Fe-based compound, which may be attributed to the smaller unit cell volume of the Co compound. Phonon modes show large explicit anharmonicity at low energies up to 5 meV. Pressure dependence of the phonon energies in the entire Brillouin zone is used to gain deeper understanding into the observed colossal thermal expansion behavior. Our analysis shows that chemical variation at the $MC_6$ octahedral (M=Co, Fe) site does not change thermal expansion behavior. It is the anharmonicity of the Ag atoms seems that drives the thermal expansion behavior. The low-energy phonon modes below 5 meV mainly involve Ag atoms. The strong anharmonic behavior of the Ag vibrations is corroborated by the pressure dependence of the phonon spectra.

**ACKNOWLEDGEMENT**

We thank Dr. Andrew L. Goodwin, University of Oxford, UK for providing $Ag_3Co(CN)_6$ for inelastic neutron scattering measurements.

TABLE I. Calculated Raman and IR frequencies (cm$^{-1}$) for Ag$_3$M(CN)$_6$ (M=Co, Fe). Irrep and Type stand for irreducible representation and type of the mode, respectively. R and I indicate whether the mode is Raman or IR active, respectively. The local point group symmetry is C$_{2h}$.

| M=Co | Calc | 27 | 38 | 40 | 46 | 48 | 48 | 71 | 84 | 85 | 131 | 133 | 139 |
|---|---|---|---|---|---|---|---|---|---|---|---|---|---|
| | Irrep | A$_u$ | B$_u$ | A$_u$ | B$_u$ | B$_u$ | A$_u$ | B$_g$ | A$_g$ | B$_g$ | A$_g$ | B$_g$ | B$_u$ |
| | Type | I | I | I | I | I | I | R | R | R | R | R | I |
| | Calc | 140 | 155 | 172 | 176 | 180 | 181 | 279 | 288 | 288 | 319 | 320 | 328 |
| | Irrep | A$_u$ | A$_g$ | A$_u$ | B$_u$ | A$_u$ | B$_u$ | B$_u$ | B$_u$ | A$_u$ | B$_g$ | A$_g$ | B$_g$ |
| | Type | I | R | I | I | I | I | I | I | I | R | R | R |
| | Calc | 419 | 424 | 428 | 462 | 464 | 468 | 502 | 503 | 507 | 554 | 554 | 559 |
| | Irrep | B$_u$ | A$_u$ | A$_u$ | A$_g$ | A$_g$ | B$_g$ | B$_u$ | A$_u$ | B$_u$ | B$_g$ | A$_g$ | A$_g$ |
| | Type | I | I | I | R | R | R | I | I | I | R | R | R |
| | Calc | 608 | 608 | 611 | 2217 | 2218 | 2220 | 2224 | 2225 | 2261 | | | |
| | Irrep | B$_u$ | A$_u$ | B$_u$ | B$_u$ | A$_u$ | B$_u$ | B$_g$ | A$_g$ | A$_g$ | | | |
| | Type | I | I | I | I | I | I | I | R | R | | | |
| M=Fe | Calc | 7 | 26 | 33 | 38 | 46 | 47 | 49 | 79 | 83 | 86 | 129 | 134 |
| | Irrep | B$_u$ | A$_u$ | B$_u$ | A$_u$ | B$_u$ | B$_u$ | A$_u$ | B$_g$ | A$_g$ | B$_g$ | A$_g$ | B$_u$ |
| | Type | I | I | I | I | I | I | I | R | R | R | R | I |
| | Calc | 136 | 153 | 172 | 173 | 178 | 179 | 277 | 286 | 288 | 317 | 319 | 327 |
| | Irrep | B$_g$+A$_u$ | A$_g$ | A$_u$ | B$_u$ | B$_u$ | A$_u$ | B$_u$ | B$_u$ | A$_u$ | A$_g$ | B$_g$ | B$_g$ |
| | Type | R+I | R | I | I | I | I | I | I | I | R | R | R |
| | Calc | 409 | 411 | 414 | 449 | 450 | 455 | 481 | 483 | 496 | 535 | 536 | 543 |
| | Irrep | B$_u$ | A$_u$ | A$_u$ | A$_g$ | A$_g$ | B$_g$ | B$_u$ | A$_u$ | B$_u$ | A$_g$ | B$_g$ | A$_g$ |
| | Type | I | I | I | R | R | R | I | I | I | R | R | R |
| | Calc | 580 | 581 | 588 | 2146 | 2148 | 2148 | 2149 | 2150 | 2183 | | | |
| | Irrep | B$_u$ | A$_u$ | B$_u$ | A$_g$ | B$_u$ | B$_g$ | A$_u$ | B$_u$ | A$_g$ | | | |
| | Type | I | I | I | I | I | I | I | R | R | | | |

FIG. 1. (Color online) The crystal structure of $Ag_3M(CN)_6$ (M=Co, Fe).

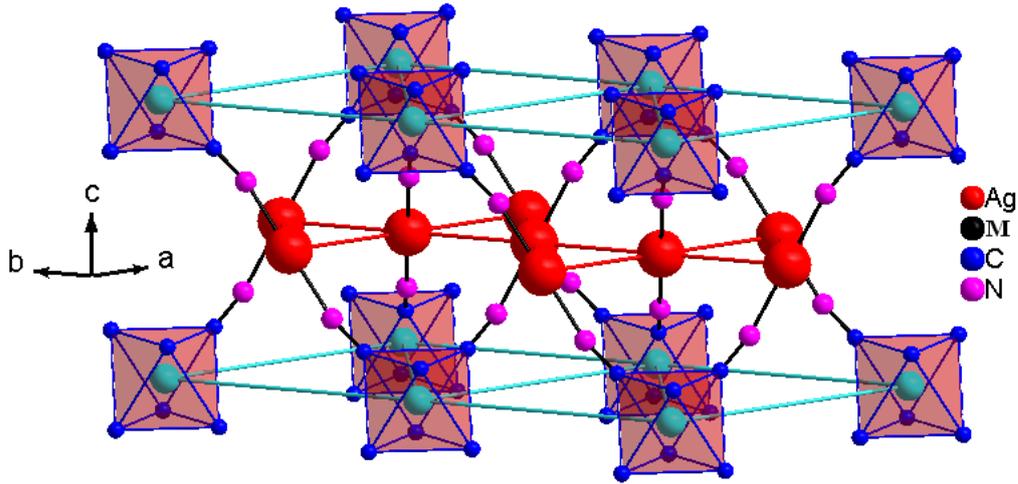

FIG. 2. (Color online) The temperature dependence of the phonon spectra for $Ag_3Co(CN)_6$. The phonon spectra are measured with an incident neutron wavelength of 4.14 Å using the IN6 spectrometer at the ILL. The calculated phonon spectra from *ab-initio* calculations are also shown. The calculated spectra have been convoluted with a Gaussian of FWHM of 10% of the energy transfer in order to describe the effect of energy resolution in the experiment.

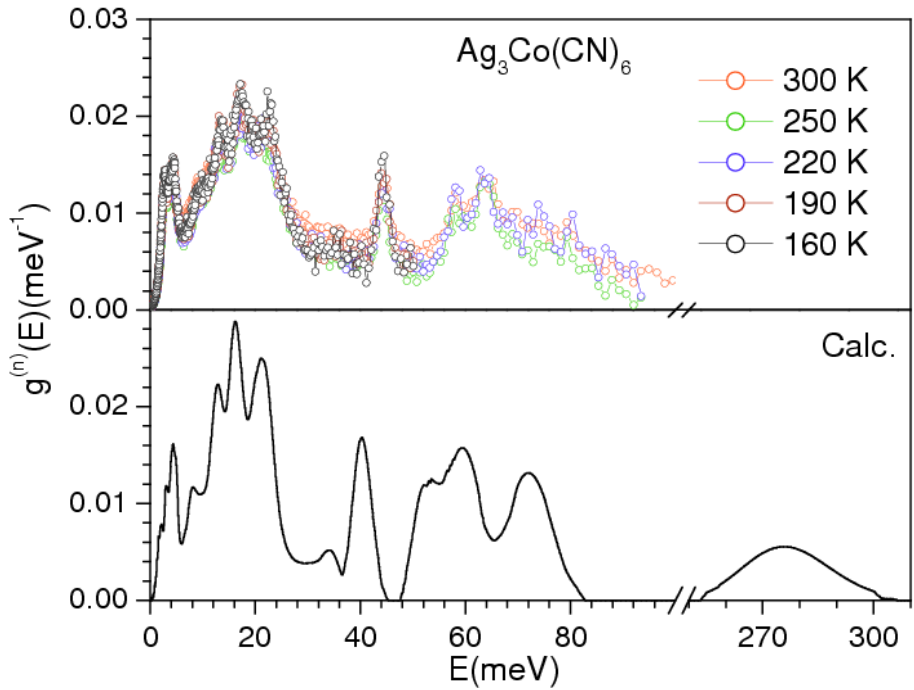

FIG. 3. (Color online) The temperature dependence of the phonon spectra for $Ag_3Fe(CN)_6$. The phonon spectra are measured with an incident neutron wavelength of 4.14 Å using the IN6 spectrometer at ILL. The calculated phonon spectra from *ab-initio* calculations are also shown. The calculated spectra have been convoluted with a Gaussian of FWHM of 10% of the energy transfer in order to describe the effect of energy resolution in the experiment.

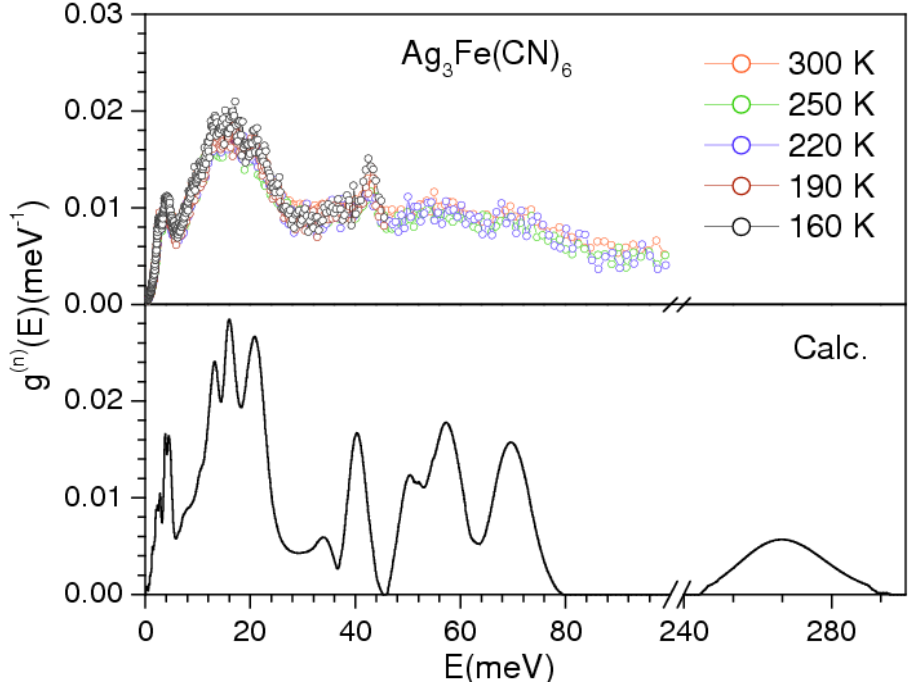

FIG. 4. (Color online) The low-energy part of the temperature dependence of the phonon spectra for $Ag_3Co(CN)_6$ and $Ag_3Fe(CN)_6$, extracted from Figures 2 and 3.

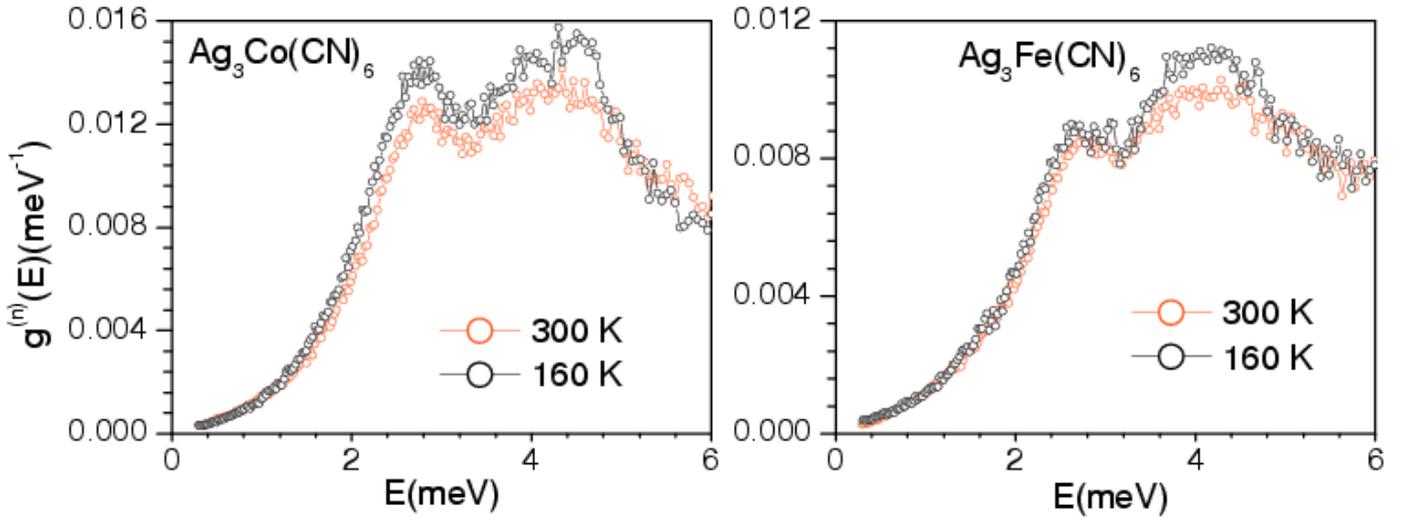

FIG. 5. (Color online) Comparison of the experimental phonon spectra for $Ag_3Co(CN)_6$ and $Ag_3Fe(CN)_6$ at 300 K.

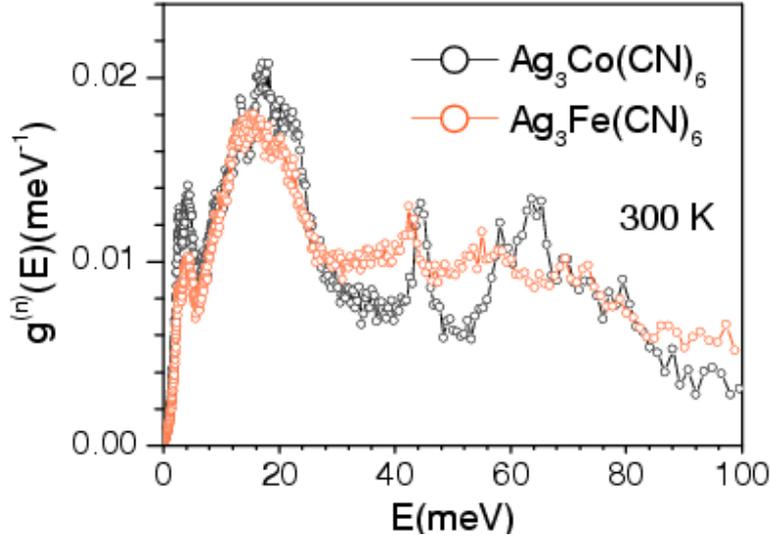

FIG. 6. (Color online) The calculated partial density of states for the various atoms in $Ag_3Co(CN)_6$ and $Ag_3Fe(CN)_6$.

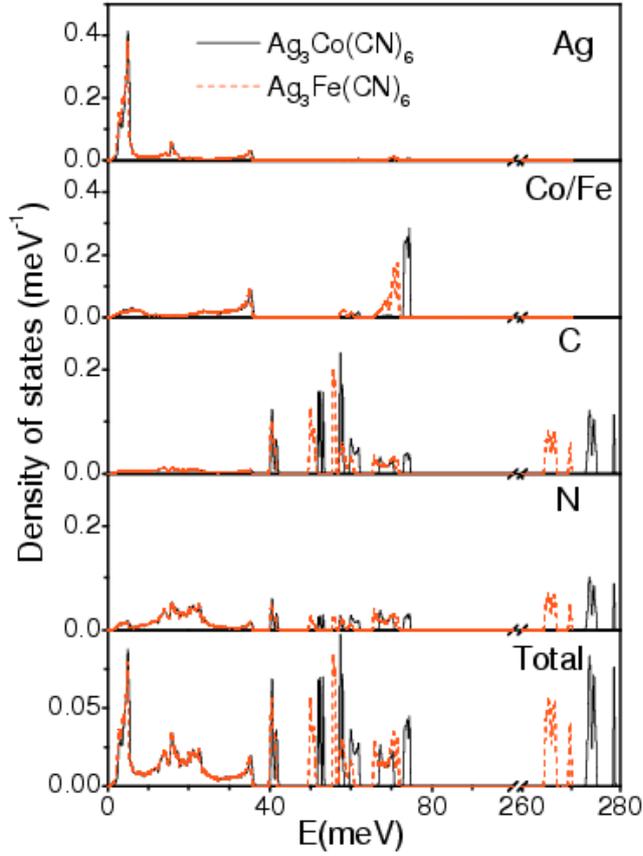

FIG. 7. (Color online) The temperature dependence of the experimental Bose-factor corrected scattering function S(Q,E) for $Ag_3Co(CN)_6$ (top panel) and $Ag_3Fe(CN)_6$ (bottom panel). For clarity, a logarithmic representation is used for the intensities. The measurements were performed with an incident neutron wavelength of 4.14 Å (4.77 meV).

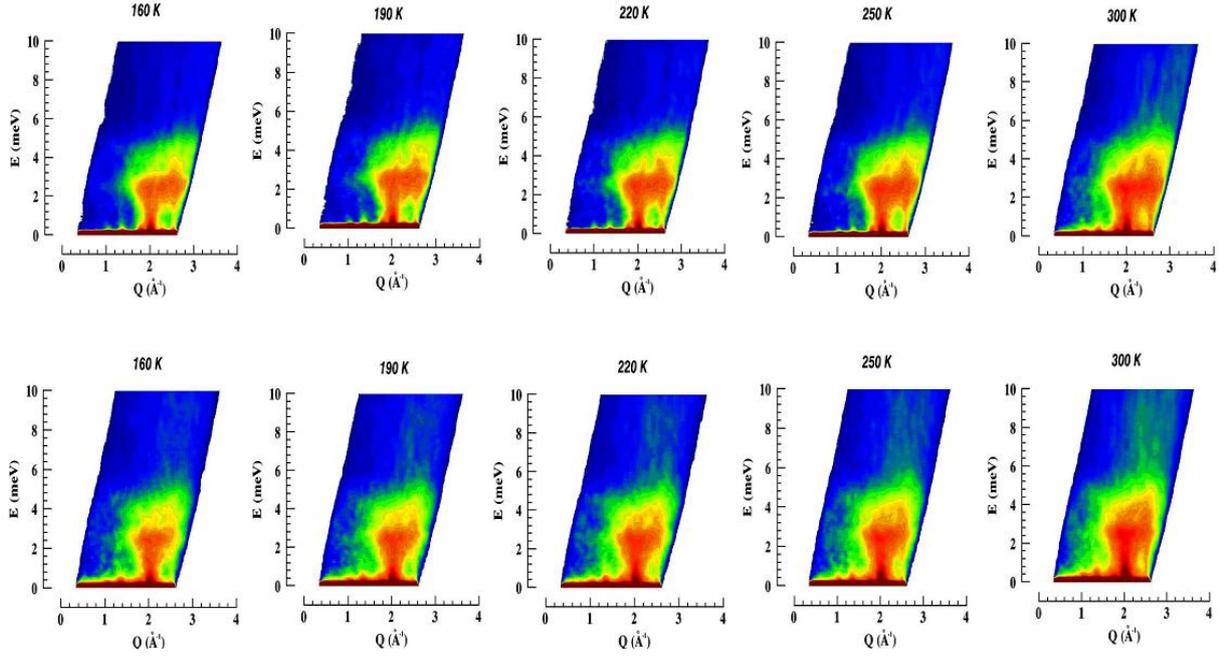

FIG. 8. The calculated phonon dispersion curves for $Ag_3Co(CN)_6$ and $Ag_3Fe(CN)_6$. The Bradley-Cracknell notation is used for the high-symmetry points along which the dispersion relations: Γ=(0,0,0), M(1/2,0,0), A(0,0,1/2), and L(0,1/2,1/2). In order to expand the *y*-scale, the six dispersion-less modes due to the cyanide stretch at about 265 to 270 meV are not shown.

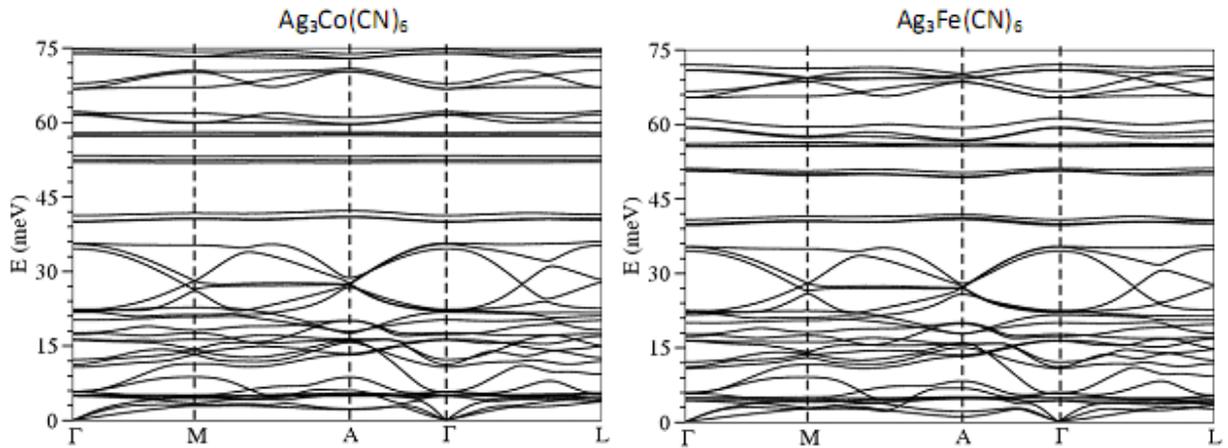

FIG. 9. The mean phonon frequencies of various atoms as obtained from ab-inito calculations in Ag$_3$Co(CN)$_6$ (full lines) and Ag$_3$Fe(CN)$_6$ ( dashed lines) and from reciprocal-space analysis of neutron diffraction data [10] in Ag$_3$Co(CN)$_6$ (symbols plus lines).

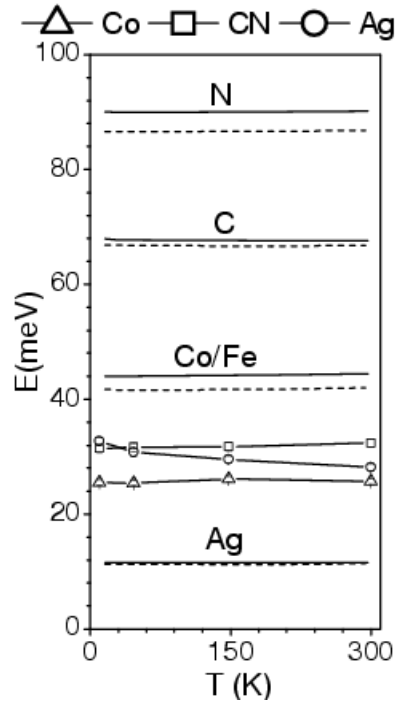

FIG. 10. (Color online) The experimental phonon spectra for Ag$_3$Co(CN)$_6$ as a function of pressure at 200 K: ambient pressure (full line), 0.3 kbar (dotted line), 1.9 kbar (dashed line), and 2.8 kbar (dash-dotted line). The measurements were done at 200 K using the IN6 spectrometer at the ILL with an incident neutron wavelength of 5.12 Å.

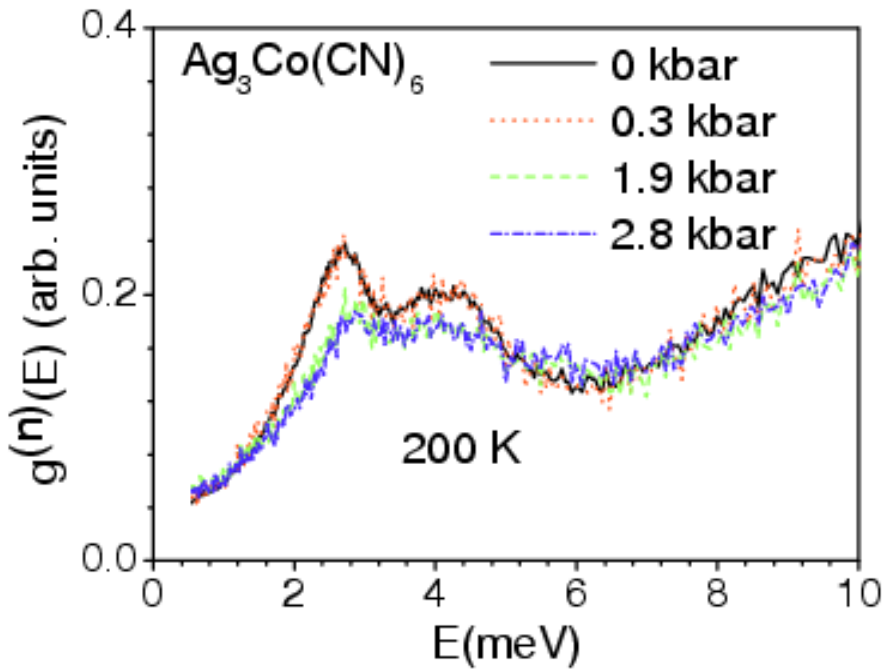

FIG. 11. The Grüneisen parameters, $\Gamma(E)$ for $Ag_3Co(CN)_6$ and $Ag_3Fe(CN)_6$ as a function of phonon energy $E$, from ab-initio calculations.

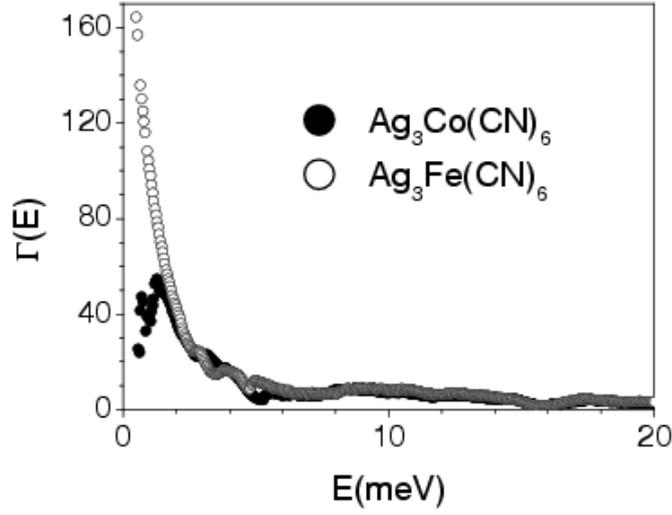

FIG. 12. The comparison between the volume thermal expansion derived from the present *ab-initio* calculations (solid line) and that obtained using X-ray diffraction [10,11] (open circles).

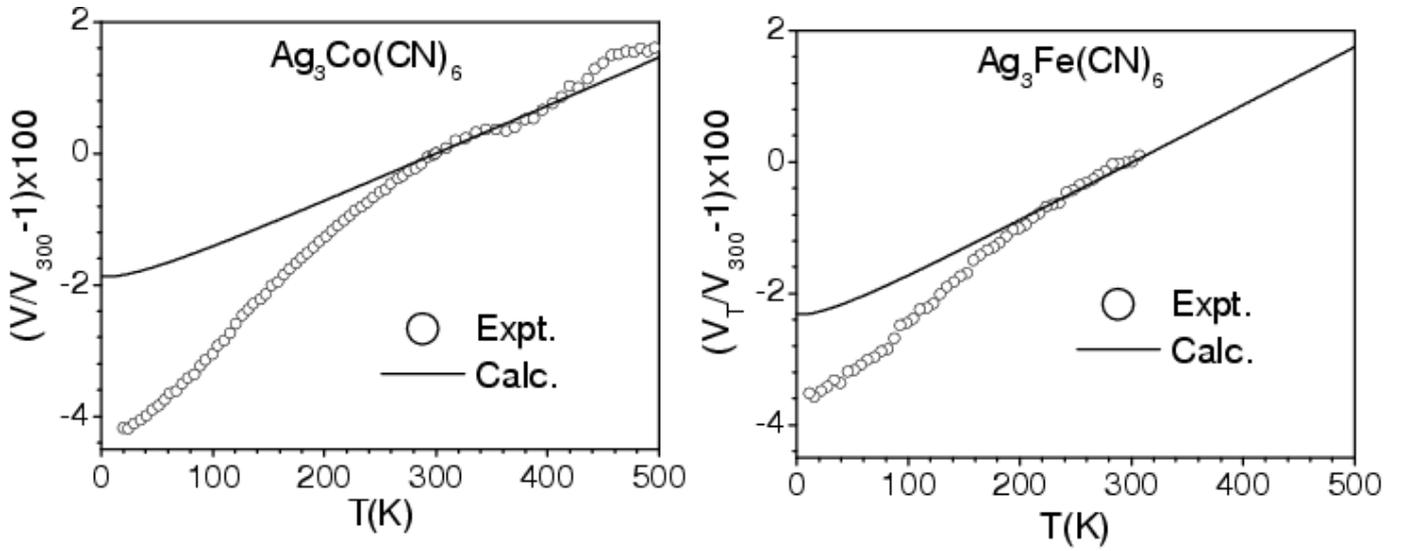

FIG. 13. The contribution of phonons of energy $E$ to the volume thermal expansion coefficient ($\alpha_V$) as a function of $E$ at 300 K in $Ag_3Co(CN)_6$ and $Ag_3Fe(CN)_6$, from ab-initio calculations.

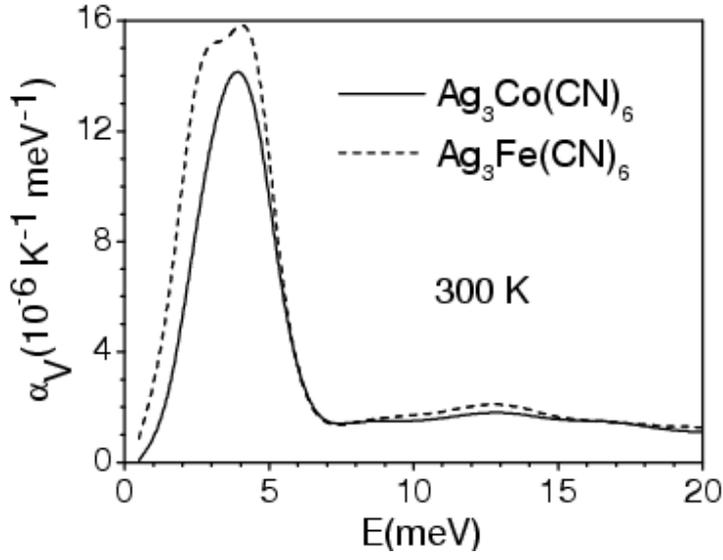

FIG. 14. (Color online) The calculated contribution to the mean squared amplitude of the various atoms arising from phonons of energy E at $T=300$ K in $Ag_3[Co(CN)_6]$ and $Ag_3[Fe(CN)_6]$, from ab-initio calculations.

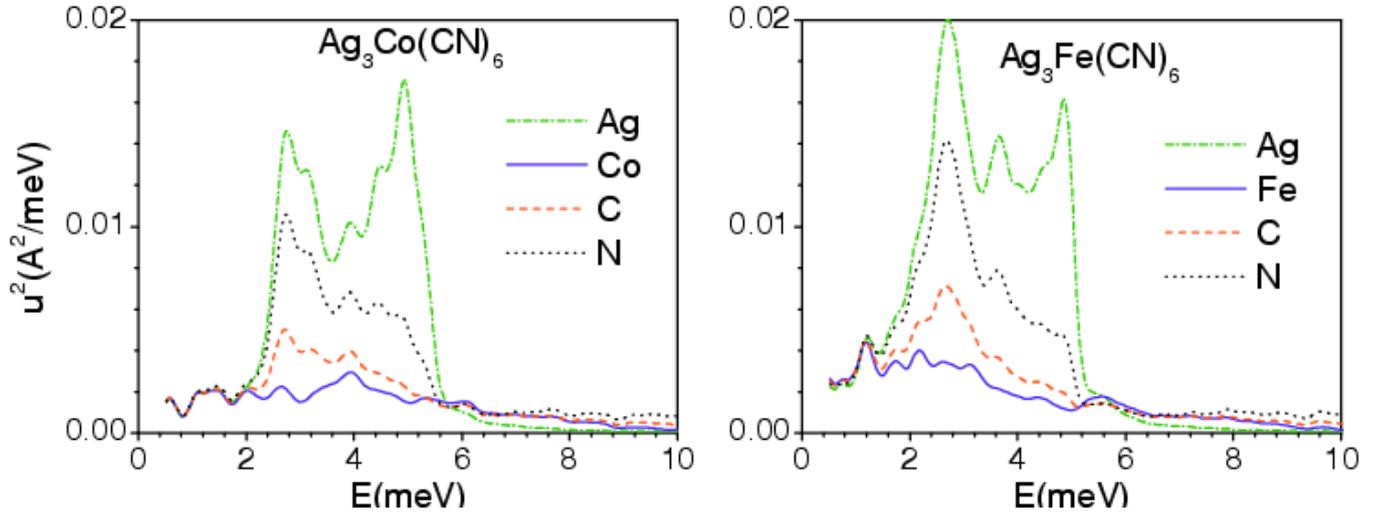